\documentclass[twocolumn,naturemag,superscriptaddress,lonbibliography]{revtex4-1}
\usepackage{natbib}
\usepackage{graphicx}
\usepackage{color}
\usepackage{amssymb}
\usepackage{amsfonts}
\usepackage{amsmath}
\usepackage{bm}
\usepackage{bbm}
\usepackage{mathtools}
\usepackage{hyperref}
\hypersetup{bookmarksnumbered=true,	unicode=false,	pdfstartview={FitH}, pdfnewwindow=true, colorlinks=true,linkcolor=blue, citecolor=blue, filecolor=blue, urlcolor=blue}

\DeclarePairedDelimiter\ket{\lvert}{\rangle}
\DeclarePairedDelimiterX\braket[2]{\langle}{\rangle}{#1 \delimsize\vert #2}

\newcommand{\ad} {\ensuremath{a^{\dagger}}}
\newcommand{\eps}{\ensuremath{\epsilon}}
\newcommand{\w} {\ensuremath{\omega}}
\newcommand{\figref}[2]{Fig. \ref{#1}{\color{blue}#2}}%
\begin{document}

\title{Parametric longitudinal coupling between a high-impedance superconducting resonator and a semiconductor quantum dot singlet-triplet spin qubit}
\author{C.~G.~L.~B\o{}ttcher}
\affiliation{Department of Physics, Harvard University, Cambridge, MA 02138, USA}
\author{S.~P.~Harvey}\thanks{\textit{Present address:} Stanford University, Stanford, CA 94305, USA}
\affiliation{Department of Physics, Harvard University, Cambridge, MA 02138, USA}
\author{S.~Fallahi}
\affiliation{Department of Physics and Astronomy, Purdue University, West Lafayette, IN 47907, USA}
\author{G.~C.~Gardner}
\affiliation{Department of Physics and Astronomy, Purdue University, West Lafayette, IN 47907, USA}
\author{M.~J.~Manfra}
\affiliation{Department of Physics and Astronomy, Purdue University, West Lafayette, IN 47907, USA}
\affiliation{School of Materials Engineering, Purdue University, West Lafayette, IN 47907, USA}
\affiliation{Birck Nanotechnology Center, Purdue University, West Lafayette, IN 47907, USA}
\affiliation{School of Electrical and Computer Engineering, Purdue University, West Lafayette, IN 47907, USA}
\author{U.~Vool}
\affiliation{Department of Physics, Harvard University, Cambridge, MA 02138, USA}
\affiliation{John Harvard Distinguished Science Fellowship, Harvard University, Cambridge, Massachusetts, 02138, USA}
\author{S.~D.~Bartlett}
\affiliation{Centre for Engineered Quantum Systems, School of Physics, The University of Sydney, Sydney, NSW 2006, Australia}
\author{A.~Yacoby}
\affiliation{Department of Physics, Harvard University, Cambridge, MA 02138, USA}

\date{\today}
\begin{abstract}
{\footnotesize 
Long-distance two-qubit coupling, mediated by a superconducting resonator, is a leading paradigm for performing entangling operations in a quantum computer based on spins in semiconducting materials. Here, we demonstrate a novel, controllable spin-photon coupling based on a longitudinal interaction between a spin qubit and a resonator. We show that coupling a singlet-triplet qubit to a high-impedance superconducting resonator can produce the desired longitudinal coupling when the qubit is driven near the resonator's frequency. We measure the energy splitting of the qubit as a function of the drive amplitude and frequency of a microwave signal applied near the resonator antinode, revealing pronounced effects close to the resonator frequency due to longitudinal coupling. By tuning the amplitude of the drive, we reach a regime with longitudinal coupling exceeding $1$ MHz. This demonstrates a new mechanism for qubit-resonator coupling, and represents a stepping stone towards producing high-fidelity two-qubit gates mediated by a superconducting resonator.
}
\end{abstract}
\maketitle 
\indent 

\section{Introduction}
Electron spins in semiconducting materials, such as gallium arsenide (GaAs) and silicon, are promising candidates for realizing a quantum computer \cite{Loss1998,Koppens2006,Ladriere2008,Kim2014,Eng2015}. Their long coherence times and fast control allow for high-fidelity single-qubit gates, reaching $\sim$99.95 \% in single-electron spin qubits \cite{Yang2019}. 
In addition to single-spin qubits, several varieties of spin qubits that are comprised of multiple spins and multiple quantum dots, including hybrid qubits, exchange-only qubits and singlet-triplet qubits ($\text{\emph{S-T}}_0$) \cite{Dohun2014,Medford2014,Wu2014}, have been demonstrated. These qubits typically have increased coupling to charge, allowing fast, voltage-controlled qubit gates.
The $\text{\emph{S-T}}_0$ qubit is desirable due to its reduced coupling to homogeneous magnetic fields and has achieved single qubit gate fidelities of 99.5 \%  \cite{Cerfontaine2020}. While two-qubit gates have previously been demonstrated for these qubits with a fidelity of $\sim$90\% \cite{Nichol2017}, these gates are slow and rely on nearest neighbor coupling, limiting scalability. Much attention is now focused on achieving long-range two-qubit coupling, for example, using arrays of quantum dots for charge transfer\cite{Fujita,Mills2019,yoneda2020coherent,Qiao2021} or a superconducting resonator by adapting circuit QED (cQED) techniques, thus making electron spins a scalable platform for quantum computing technology.

Extensive work on implementation of cQED techniques in spin qubits has recently been demonstrated \cite{Samkharadze1123,Landig2018,Borjans2020,Mi2017,Mi20172}, and despite promising progress \cite{Borjans20202}, a two-qubit gate has not yet been achieved. The qubit-resonator coupling explored relies on the strong electric fields produced by a resonator, which couple to the dipole moment of a spin qubit. \\
The most commonly considered coupling scheme is a transverse coupling between the spin and resonator, where an excitation of the spin qubit can be exchanged for a resonator excitation\cite{Wallraff2004}. This requires the qubit energy splitting to be near the resonator frequency, and typically leads to lower lifetimes due to the Purcell effect. In recent years, there has therefore been growing interest in alternative coupling schemes based on longitudinal interactions, which do not have these limitations \cite{Blais2004,Kerman2013,Schuetz2017,Billangeon2015,Didier2015,Royer2017,Ruskov2021}.
\begin{figure}[t]
	\includegraphics[width= 3.4 in]{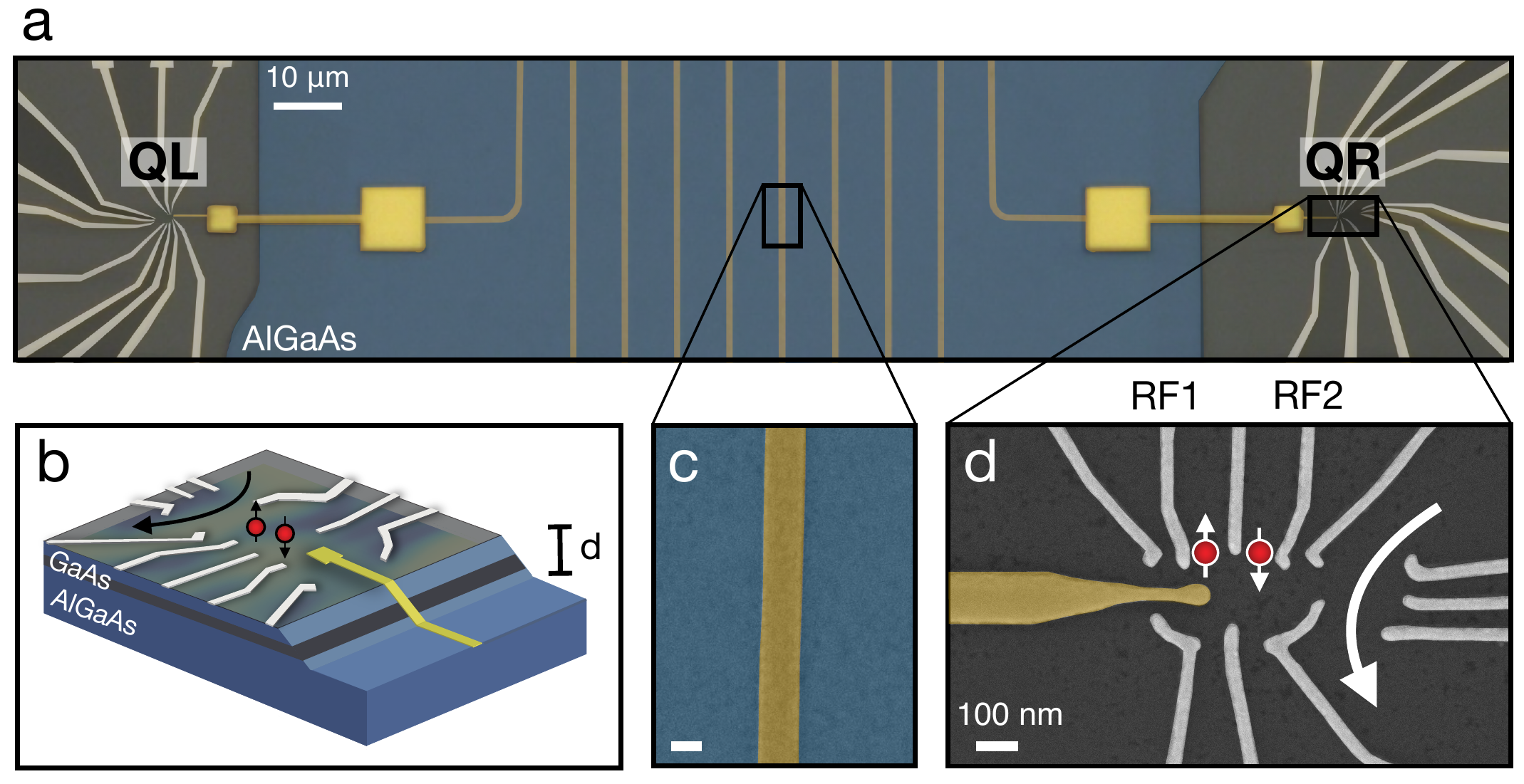}
	\caption{\textbf{Device layout}. \textbf{a)} False colored scanning electron microscope (SEM) image of two DQDs that are placed at each end of a superconducting resonator made from thin film NbN. The resonator is meandered across the etched part of the sample where 2DEG has been removed. \textbf{b)} Illustration of the resonator climbing the edge of the etched region with a height of $d \sim90~\text{nm}$. It couples capacitively to the DQD through the electric field of the resonator. \textbf{c)} To maximize the coupling, the resonator impedance can be enhanced by NbN, a material with large kinetic inductance, and reducing the center conductor width to 150 nm. \textbf{d)} SEM image showing right DQD. Each DQD requires a set of DC gates to define the quantum dots and a set of RF gates (RF1 and RF2) to have fast control of the $\text{S-T}_0$ energy splitting.}
	\label{fig1}
\end{figure}
Spin qubits are highly amenable to longitudinal coupling, although it has not been demonstrated experimentally before. In previous theoretical work \cite{Harvey2018}, such a coupling scheme was explored for singlet-triplet qubits, predicting encouraging average two-qubit gate fidelities of 96\% and gate times of the order of 10 ns. 
This approach, analogous to the M\o{}lmer-S\o{}rensen gate \cite{Molmer1999} that is commonly used for high fidelity two-qubit gates in ion trap qubits \cite{Kirchmair2009,Haffner2008}, relies on a purely longitudinal interaction between the spin and resonator to produce a two-qubit coupling.

In this article, we demonstrate experimental efforts towards achieving longitudinal coupling between a singlet-triplet ($\text{\emph{S-T}}_0$) qubit and high-impedance superconducting resonator. We show that our device has significant longitudinal coupling, tunable by a direct drive, in addition to a fixed spurious dispersive coupling. We present a measurement sequence that allows one to separate each coupling term and measure their individual coupling strengths. The sequence takes advantage of the qubit's exquisite sensitivity, enabling us to extract resonator parameters as well as qubit-resonator coupling strengths.
By tuning the drive amplitude we can achieve strong longitudinal coupling, which is an exciting new regime within hybrid circuit QED systems as well as an important stepping stone towards producing two-qubit coupling mediated by a resonator. 

\section{Design of the device} 
The device consists of two double quantum dots (DQDs) formed in Si-doped GaAs/AlGaAs heterostructure with a two-dimensional electron gas (2DEG) located $\sim$90 nm below the surface. The two DQDs are coupled to a high-impedance superconducting resonator and separated by $100~\mu$m as illustrated in \figref{fig1}{a}. The 2DEG is removed using chemical etching in a large area, leaving only a separate island (mesa) for each DQD. The two spatially separated DQDs are each tuned to be $\text{\emph{S-T}}_0$ qubits, and due to their large distance from one another, the only coupling between them is mediated by a superconducting resonator. The resonator climbs the mesa and capacitively couples to the left and right DQD (\figref{fig1}{b}), marked QL and QR in \figref{fig1}{a}. The resonator is fabricated in the etched area from a 20 nm superconducting film made of niobium nitride (NbN) and meandered across the sample. Using a thin film of NbN as the resonator material, one can obtain a large kinetic inductance, $L_K$. The kinetic inductance, $L_K = (m_e/2n_se^2)(l/A)$ \cite{Tinkham}, depends on the superfluid density, $n_s$, and scales with resonator length $l$ and cross-sectional area $A$, thus we achieve a high impedance close to $Z_r=\sqrt{(L_K+L_m)/C_r}\sim 2~\text{k}\Omega$ for a resonator design with a meander width of 150 nm (\figref{fig1}{c}). The small dimensions minimizes resonator capacitance $C_r$, and magnetic inductance $L_m$, such that the resonator is largely dominated by its kinetic inductance. The resonator’s high impedance makes it well suited for coupling to systems such as electrons in DQDs, which have small electric dipole moments. 

Our $\text{\emph{S-T}}_0$ qubits each consist of two electrons trapped in a DQD defined using electrostatic gates for static potential confinement shown in \figref{fig1}{d}. The logical subspace of the qubits consists of the singlet, $|S \rangle = (|\uparrow\downarrow \rangle-|\downarrow\uparrow  \rangle)/\sqrt{2}$  and triplet  $|T_0 \rangle = (|\uparrow\downarrow \rangle+|\downarrow\uparrow  \rangle)/\sqrt{2}$, states. We apply a static in-plane magnetic field of $\sim700~\text{mT}$ making higher energy states energetically inaccessible. The energy splitting $J(\eps)$, splits $S$ from $T_0$ and is tuned on a nanosecond timescale by the difference in chemical potential $\eps$, set by the two radio frequency (RF) gates, labeled RF1 and RF2, that enable fast pulsing and control. 

Readout in our device is different from what is typically done in circuit QED experiments.  We do not include a direct feedline port to control and read out the resonator, simplifying the design and improving the resonator quality factor and coupling.  Rather, the qubits are measured using a sensor dot proximal to each DQD to measure the charge state of the DQD. The resonator is excited using the DQD gates that are capacitively coupled to it and described in the next section. 

\section{Probing the coupled resonator-qubit system} 
We introduce a measurement technique, based on a Hahn-echo-like sequence, to characterize the qubit-resonator interaction. Due to its increased coherence time, the Hahn-echo pulse sequence offers greater sensitivity than a typical Ramsey experiment and can be used, for instance, to characterize the noise environment seen by the qubit \cite{Taylor2008,Medford2012,Dial2013}. It can also be used to measure changes in the exchange splitting of the qubit, $J(\epsilon)$, between the first and second half of the pulse sequence, which we use to extract changes in the electrostatic environment of the qubit.

The electrostatic environment of each qubit is determined by the control lines together with the resonator.  We can control each qubit by applying voltages to the nearby RF gates, with which we apply a static potential $\epsilon_0$ and a direct RF drive $\epsilon_d \cos \omega_d t$, where $\omega_d$ is the drive frequency. We will use superscripts L,R on the drive amplitude ($\epsilon^{L,R}_d$) to indicate if the drive is sent to the left or right qubit. In addition, the qubit is sensitive to the voltage fluctuations of the resonator $V_r =V_0(a+a^\dagger)$ where $a$ is the resonator annihilation operator and $V_0 = \sqrt{\hbar Z_r/2}\omega_r$ is the zero-point voltage fluctuation, set by the frequency of the resonator $\omega_r$ and its impedance $Z_r$.
The chemical potential at each quantum dot can thus be expressed as $\epsilon = \epsilon_0 +\eps_d \cos\omega_d t+ec_rV_r$, where $c_r$ is the lever-arm of the resonator fluctuations on the qubit.  Moving to an interaction picture relative to the drive frequency $\omega_d$, the resulting qubit energy splitting $J(\epsilon)$ leads to the qubit-resonator Hamiltonian (see Ref.~\cite{Harvey2018}) given by 
\begin{align} \label{eq:H}
H_{\rm int} &=\hbar \Delta \ad a +J(\epsilon_0)\sigma_z+ \tfrac{1}{2}g(a+\ad)\sigma_z  + \tfrac{1}{2}\chi \ad a \sigma_z, 
\end{align}
where $ \Delta = \w_r - \w_d$ is the detuning, and $g =\frac{1}{2} \frac{d^2J}{d\eps^2}\bigr|_{\eps_0} c_r V_0 \epsilon_d$ and $\chi =  \frac{d^2J}{d\eps^2}\bigr|_{\eps_0} c_r^2 V_0^2 $ are the two coupling strengths.
The Hamiltonian can be written in the simplified form $H = \hbar \Delta \ad a + \tilde{J}\sigma_z$ by defining the modified qubit energy splitting $\tilde{J} = J(\epsilon_0)+\tfrac{1}{2}g(a+\ad)+ \tfrac{1}{2}\chi\ad a$, where the last two terms represent the qubit interaction with the resonator.  We refer to the first term, proportional to $g$, as the \emph{longitudinal term}, and note that it can be tuned by the amplitude of the drive, $\epsilon_d$.  The second term, proportional to $\chi$, which we call the \emph{dispersive term}, is independent of the drive. Following Ref.~\cite{Harvey2018}, we are ultimately interested in implementing a two-qubit coupling in the case where $c_rV_0/\epsilon_d\ll 1$, i.e., where the dominant interaction is longitudinal and set by coupling strength $g$.  However, to fully characterize our device, we wish to determine each qubit's individual coupling strengths $g$ and $\chi$ to the resonator. We therefore focus primarily on the interaction of a single qubit with the resonator. Our device architecture allows us to independently measure each qubit and control the two coupling terms, simply by driving the resonator using near and far qubit RF gates with respect to the active qubit, as we now describe.

 \begin{figure}[t]
	\includegraphics[width= 3.5 in]{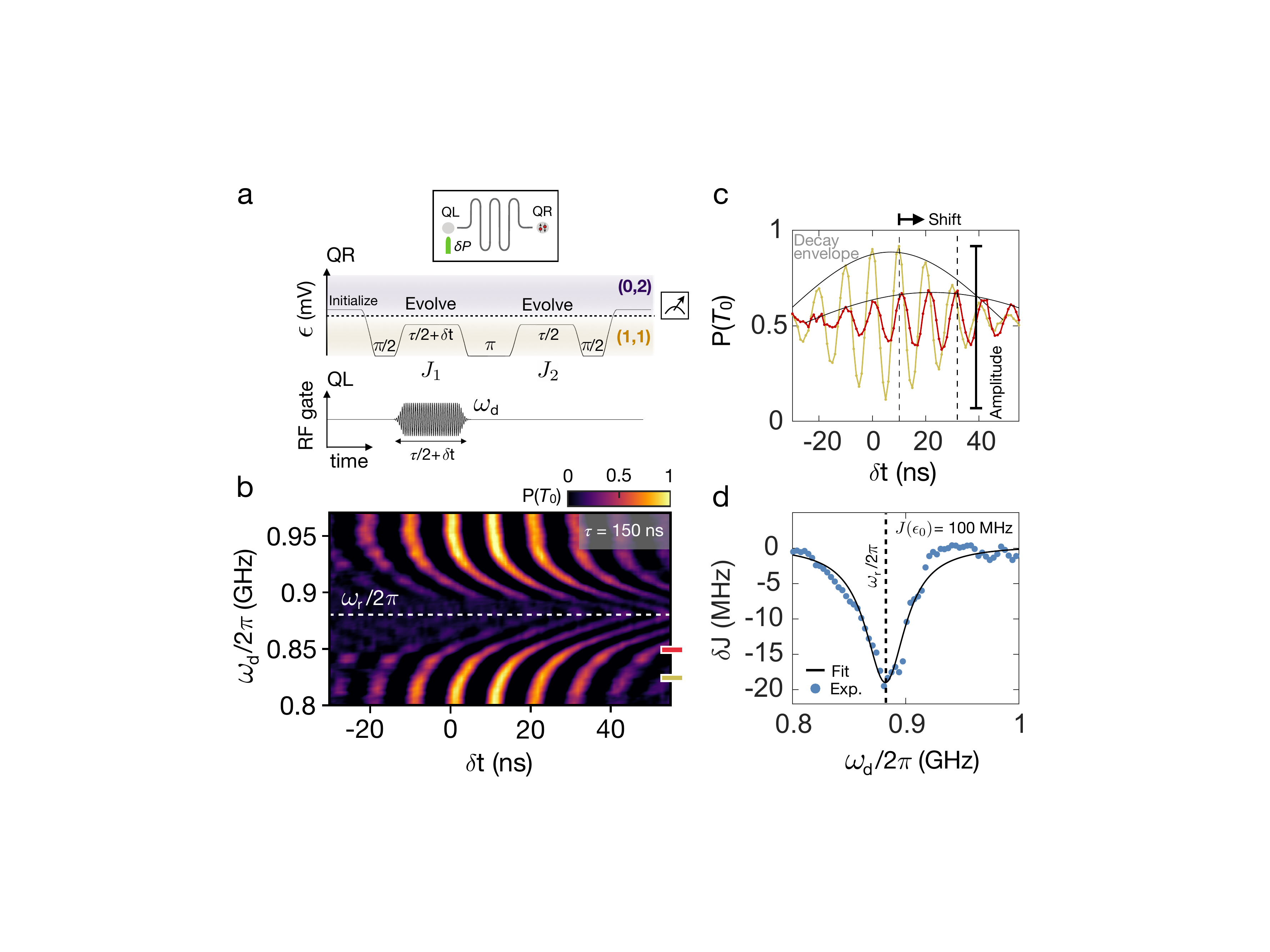}
	\caption{\textbf{Measurement sequence for probing resonator-qubit coupling}. \textbf{a)} Echo pulse sequence used to measure the exchange splitting of QR while driving the resonator using the RF gate of QL (dark green in cartoon inset). \textbf{b)} Exchange oscillations measured at several drive frequencies, indicating the qubit-resonator interaction near the resonator frequency $\omega_r$. \textbf{c)} Two line cuts at different values of $\omega_d$ reveal a phase shift and rapid qubit dephasing when the drive frequency is close to the resonator frequency $\omega_r/2\pi = 0.88~\text{GHz}$. The decay envelopes fit to a Gaussian decay function given by $T^*_2$. \textbf{d)} Extracted values of $\delta J$ as a function of $\omega_d$ fit to a Lorentzian function to extract resonator decay rate $\kappa$, resonance frequency $\omega_r$ and coupling strength $\chi$. 
			}
	\label{fig2}
\end{figure}
We first consider a measurement that allows us to characterize the dispersive term $\chi$.  To measure $\tilde{J}$, the standard Hahn-echo sequence is modified by introducing a drive of the resonator during the first evolution time $\tau/2$ of QR as illustrated in \figref{fig2}{a}. As indicated in the device sketch, we use an RF signal with power $\delta P$ to turn on the gate marked in green on the left side of the resonator, the opposite antinode of QR. The left qubit, QL, is kept far detuned, and it can effectively be ignored for the remainder of the paper. Because there is no direct drive of QR in this experiment, only coupling through the resonator, we expect $g=0$ and that this sequence will generate a coupling given by only the dispersive term with strength $\chi$.  
As the frequency $\omega_d$ of the RF pulse is tuned near the fundamental frequency of the resonator $\omega_r$, the resonator will be excited and interact with the qubit. The qubit's energy splitting is modified by the qubit-resonator interaction to be $J_1$, which differs from the splitting in the second half of the sequence where RF excitation is turned off, $J_2$.  
The echo measurements reveal a decay envelope, captured by sweeping the length of the first pulse by small increments of $\delta t$. The maximum amplitude reveals the extent to which the state has dephased, while the shape of the envelope and width arise from an effective single-qubit rotation for a time $\delta t$ and envelope associated with $T^*_2$ \cite{Dial2013}. In \figref{fig2}{b} we plot the exchange oscillations measured for a set of drive frequencies and observe a significant phase shift of the decay envelope when $\omega_d\sim \omega_r$. The shape is well-captured by a Gaussian decay function similar to Ref.~\cite{Dial2013} with $T^*_2\sim 250~\text{ns}$, presented in \figref{fig2}{c} for two line cuts, one taken far off resonance at $0.82~\text{GHz}$ and one taken close to resonance at $0.85~\text{GHz}$. There are two clear effects. First, a significant shift of the envelope close to resonance suggests an accumulated phase during the RF on-sequence due to the qubit's interaction with the resonator. The magnitude of the phase shift is given by $\theta = (J_1-J_2)\tau/2=\delta J\,\tau/2$. We fit the oscillations and plot $\delta J$ as a function of drive frequency in \figref{fig2}{d}, which is well described by a Lorentzian function.  We extract the resonator parameters $\omega_r/2\pi = 0.88~\text{GHz}$ and $\kappa/2\pi \approx 50~\text{MHz}$ corresponding to a $Q=\omega/\kappa \approx 20$ discussed later, as well as the dispersive coupling strength $\chi/2\pi \approx 0.2~\text{MHz}$ for a qubit detuning leading to $J(\eps_0)\approx 100~\text{MHz}$ and fixed drive amplitude.  The data show an additional effect: a significant suppression of the oscillation amplitude closer to resonance $\w_r$ (\figref{fig2}{c}), suggesting a rapid dephasing of the qubit.  We return to this effect in Sec.~V. 

\section{Longitudinal coupling} 
\begin{figure}[]
	\includegraphics[width= 2.6 in]{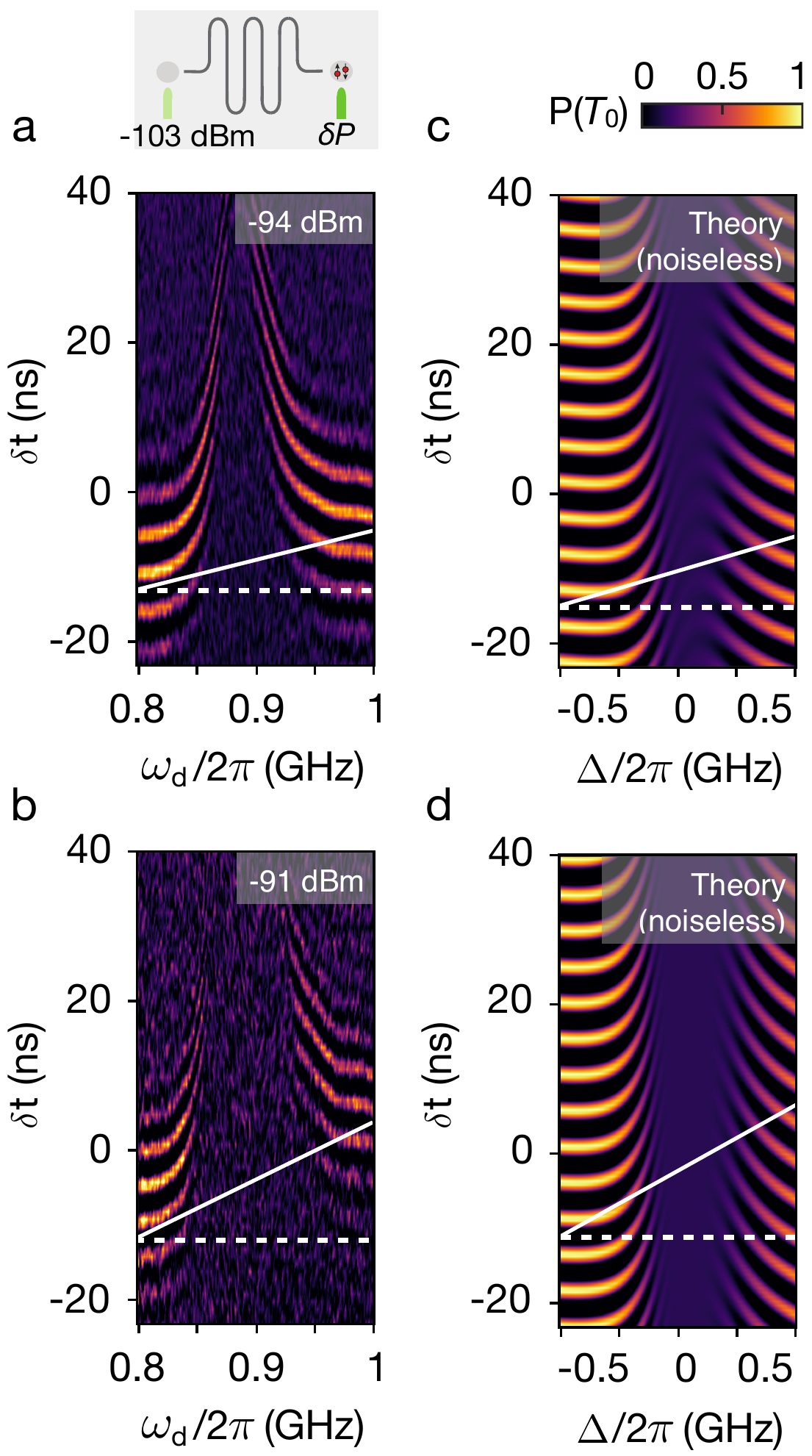}
	\caption{\textbf{Exchange oscillations in double-driven regime}. Inset: We drive the resonator by sending a RF signal to the gate placed at its right antinode (dark green) with power $\delta P$ and simultaneously sending a signal to the gate at the left antinode (light green), the opposite side of the active qubit, which is kept at a fixed power. \textbf{a)} Exchange oscillations when $\delta P=-94 ~\text{dBm}$ reveal a large phase shift near $\omega_r$ marked with the white dashed line. The asymmetry around resonance is marked by a white line tilted with respect to a horizontal reference line (white dashed line). \textbf{b)} Power sent to the right gate is increased to $\delta P = -91~\text{dBm}$ leading to a larger phase shift and an enhanced asymmetry. \textbf{c)} Solutions to the master equation for the coupled qubit-resonator system produce a similar asymmetry to (a)  when the $g$ term is the dominant coupling term in Eq.~1. \textbf{d)} By further increasing the strength of $g$, the simulations reproduce the enhanced asymmetry in (b), corresponding to the highest drive power.    Simulations do not include an explicit dephasing term for the qubit.    
	}
	\label{fig3}
\end{figure}

Having characterized the drive-independent dispersive coupling term, we now turn our focus to the longitudinal term that would form the basis for a two-qubit coupling scheme.  To generate a longitudinal coupling with coupling strength $g$, a second drive is introduced to simultaneously modulate the qubit, QR, at frequency $\omega_d$ with a tunable drive power $\delta P$, while keeping the left drive power fixed (illustrated by the device sketch in \figref{fig3}{}). Drive frequency of left and right drive are swept simultaneously, thus $\omega_d$ represents both drive frequencies.  By systematically increasing the right drive power, $\delta P$, we can study the competition between the longitudinal and dispersive coupling terms, $g$ and $\chi$.

In \figref{fig3}{a},{\color{blue}b} exchange oscillations for two values of $\delta P$ are observed, similar to \figref{fig2}{a}. The oscillations reveal an asymmetry in the shift of the echo envelope around the resonance frequency that is enhanced with larger drive power.  Again, we observe that qubit dephasing on resonance increases with power; we return to this effect in the next section.

By solving the master equation of the coupled system described by the Hamiltonian in Eq.~1, an asymmetry similar to the one seen in the data is found; see \figref{fig3}{c},{\color{blue}d}. The system is solved by including damping from the resonator, $\kappa/2\pi = 50 ~\text{MHz}$, and tuning the ratio between the two coupling terms. This reveals that when $g$ is large, the asymmetry is pronounced, suggesting this effect is entirely due to the longitudinal term in Eq.~1.  

This measurement allows us to investigate the competition between the longitudinal and dispersive terms.  We can understand the qubit-resonator interaction by using a simple semi-classical description of the state of the resonator as a coherent state with complex amplitude $\alpha = |\alpha|e^{i\theta}=\frac{\eps_d}{\Delta + i\kappa/2}$.  With the resonator state described this way, the qubit energy splitting $\tilde{J}$ can be expressed as
\begin{equation}
  \tilde{J} = J(\eps_0) + g \text{Re}(\alpha) + \chi |\alpha|^2\,.
\end{equation}
The second term on the right side is proportional to the real part of $\alpha$ and will therefore change sign with detuning, while the last term, which is proportional to the square of the amplitude, will remain positive for all values of detuning.  This behavior explains the asymmetry of the data shown in \figref{fig3}{a},{\color{blue}b}. 

\begin{figure}[t]
	\includegraphics[width= 3.4 in]{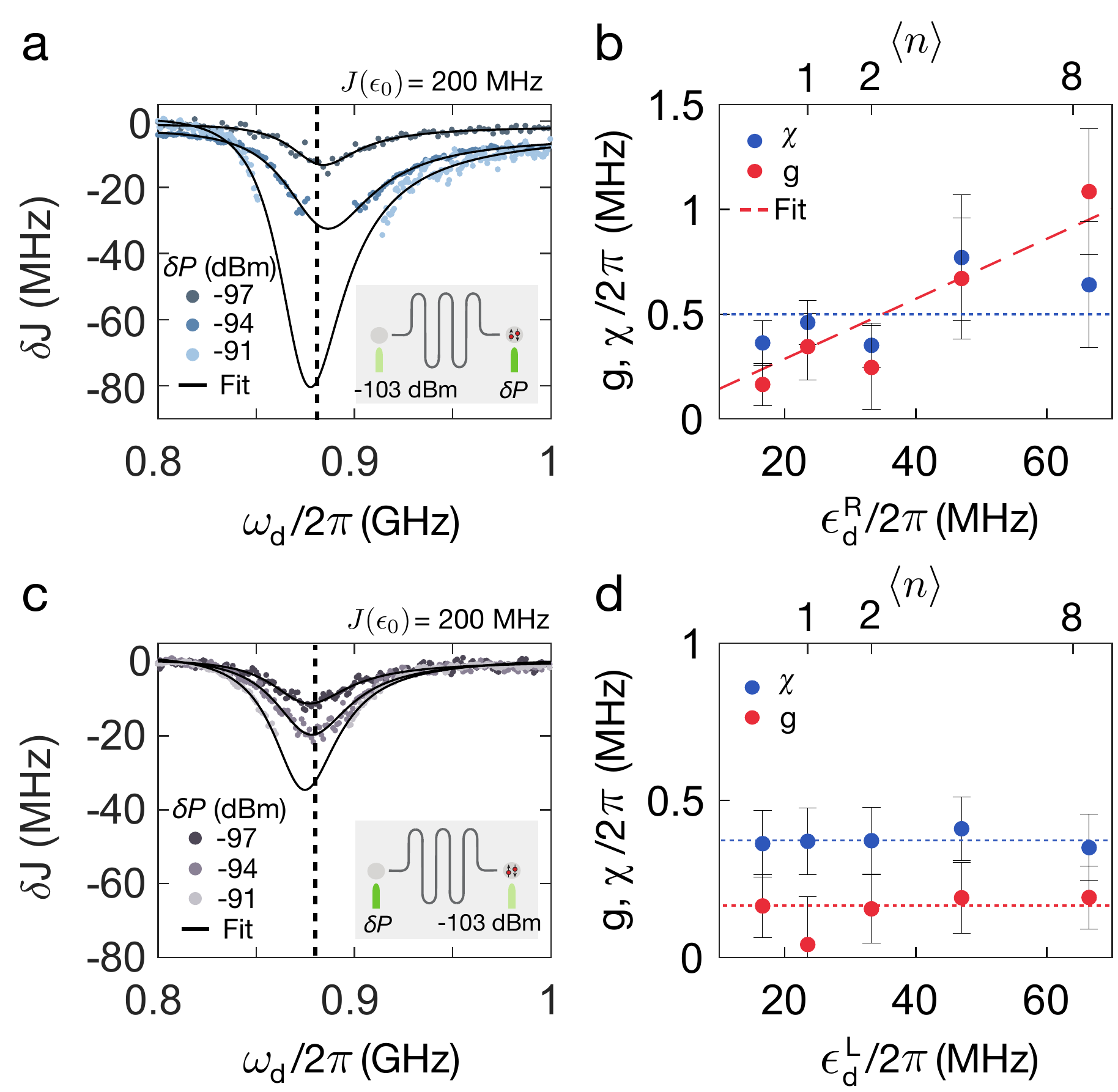}
	\caption{\textbf{Parametric longitudinal coupling}. \textbf{a)}
	Extracted $\delta J$ as a function of $\omega_d$ when right drive power is varied (dark green) while left drive power is kept constant (light green), as indicated in figure inset. The asymmetry around the resonator frequency $\omega_r$ (black dashed line) is greatly enhanced with drive power. We fit our model, Eq.~2, to extract coupling strengths $g$ and $\chi$, presented in \textbf{b)}, as a function of drive power $\epsilon_d$ (bottom horizontal axis) or of average photon number $\langle n\rangle  = (2\epsilon_d/\kappa)^2$ (top horizontal axis).  The coupling strength $g$ is observed to be tunable with linear dependence on the right drive amplitude (a square-root dependence on average photon number), while $\chi$ fluctuates around a constant value. At high power $g$ exceeds $\chi$, thus providing the dominant coupling to the resonator. \textbf{c)} Figure inset: left drive power is varied, keeping right drive fixed. Extracted $\delta J$ as a function of $\omega_d$ show a very symmetric resonance and fits to the model lead to the extracted coupling strengths presented in \textbf{d)}. Fits to our model show that both coupling strengths, $g$ and $\chi$, are independent of the left drive power.
	}
	\label{fig4}
\end{figure}

Using this simple model, we fit the extracted phase shifts, $\delta J$, from \figref{fig3}{a},{\color{blue}b} to Eq.~2 for several values of the drive power, as shown in \figref{fig4}{a}.  We observe good agreement between the model and data, and accurately capture the asymmetry around resonance. The coupling strengths are fit parameters in the model and presented in \figref{fig4}{b} as a function of the right qubit drive power $\epsilon^R_d$.  We can express the coupling parameters in terms of the average number of photons $\langle n \rangle$ generated by the drive power from the relation $\eps_d = \sqrt{\langle n\rangle}\kappa/2$. From the definitions of the two coupling coefficients, $g$ and $\chi$, we expect $\chi$ to be independent of drive power while $g$ should increase linearly with drive power.  Indeed, we find $\chi/2\pi \approx 0.5 ~\text{MHz}$ is constant up to small variations, and therefore independent of drive power, as shown in \figref{fig4}{b}.  For $g$, we observe $g/2\pi \approx 0.15 ~\text{MHz}$ at lowest value of drive power and $g/2\pi \approx 1 ~\text{MHz}$ at the highest value. The expected increase is linear with $\eps_d^R$. We note the larger error bars in the high drive power regime and that $\chi$ fluctuate more.  

To further explore the competition between $g$ and $\chi$, the left drive is now varied while the right drive is fixed. The result is presented in \figref{fig4}{c}, and we note that there is no asymmetry apparent around resonance even at the highest power. In this regime, $g/\chi \approx 0.3$, so that $\chi$ is the dominant signal. 
The fitting procedure to Eq.~2 is repeated for this coupling regime and coupling strengths are extracted and presented in \figref{fig4}{d} as a function of $\eps_d^L$. Both coupling strengths are found to be constant with small variations in this range of power. The average value for $\chi/2\pi \approx  0.4 ~\text{MHz}$ is consistent with the results shown in \figref{fig4}{b} and $g/2\pi \approx 0.15 ~\text{MHz}$ is similar to the lowest value of $g$ observed previously.
  
Driving the resonator using near and far RF gates with respect to the active qubit allows us to separate the contributions from the two coupling terms.  We can then consider tuning to the regime such that the dominant effect is longitudinal, with $g/2\pi \approx 1~\text{MHz}$, which is done by turning on the maximum drive power. Due to direct coupling between the RF gate and qubit, we are limited in further increasing the drive power as it will interfere with state preparation of our qubit. Other ways to enhance coupling strength are to increase the impedance of the resonator, thus enhancing the voltage fluctuations at the antinode, and to increase the lever arm, $c_r$.
Lastly, we note that both coupling strengths also depend on $J(\eps)$. Using the empirical approximation $J(\eps_0)\propto J_0\exp(\eps/\eps_0)$, and hence $\chi, g \propto \frac{d^2J}{d\eps^2}\bigr|_{\eps_0} \sim J/\eps^2_0 $. As a result the couplings will scale linearly with the $J$, so  it seems advantageous to increase $J$ as much as possible. 
The expected scaling with $J$ can be observed by comparing the single drive result in \figref{fig2}{d} with the results shown in \figref{fig4}{}. \figref{fig2}{} was produced with a qubit detuning of $J(\eps_0)\approx 100~\text{MHz}$ leading to $\chi/2\pi \approx 0.20 ~\text{MHz}$, while \figref{fig4}{} was produced with $J(\eps_0)\approx 200~\text{MHz}$, leading to $\chi/2\pi \approx 0.40 ~\text{MHz}$, consistent with linear scaling. This approach may not improve gate fidelities because qubit noise increases with a similar scaling \cite{Dial2013}. Instead, this is an optimization problem discussed in Ref.~\cite{Harvey2018}, where an optimal drive is found by considering the type of qubit noise measured for S-T$_0$ qubits in GaAs \cite{Dial2013}, leading to the maximum two-qubit gate fidelity for this system.

\section{Measurement-induced dephasing} 
Finally, we turn to the rapid dephasing of the qubit when the drive frequency is tuned close to that of the resonator, $\omega_d\approx\omega_r$.  This provides us with an additional tool with which to study the interaction between qubit and resonator.  \figref{fig3}{} shows that the dephasing rate is enhanced with drive and the width is of the order $\kappa$.
Such effects have previously been explored by looking at the resonator back-action on the qubit, an effect known as measurement-induced dephasing \cite{Clerk2010, Gambetta2006}.  

Formally, this requires studying photon noise, or fluctuations in the photon number operator $n = a^\dagger a$, which couples to $\sigma_z$ with coupling strength $\chi$, reflecting noise through the dispersive coupling to the resonator \cite{Gambetta2006}. However, since there are both dispersive and longitudinal coupling terms, fluctuations in the operator $x = a+ a^\dagger$ that couples to $\sigma_z$ with a coupling strength $g$ must also be considered. In both cases, the limit $g,\chi \ll \kappa$, so the dephasing can be derived using Fermi's Golden Rule, and is written as follows \cite{Ithier2005}: 

\begin{align}
    \Gamma_\phi &= \Gamma^\chi_\phi+\Gamma^g_\phi+\Gamma^{g\chi}_\phi \nonumber \\ 
  &= \frac{\chi^2}{2}S^\chi_{nn}(\omega=0)+\frac{g^2}{2}S^g_{xx}(\omega=0)+g\chi S^{g\chi}_{xn,nx}(\omega=0)
\end{align}

The first two terms arise from pure $\chi$ and $g$ effects, respectively, and the third as a result of their cross coupling. $\Gamma_{\phi}$ can be derived as a function of detuning by calculating the power spectral densities (described in Methods), $S_{AB}(\omega)=\int_{-\infty}^{\infty}dt \, C_{AB}(t) \, e^{-i\omega t}$ and the Fourier transform of the correlation functions, $C_{AB}(t) = \langle A(t)B(0)\rangle - \langle A(t)\rangle\langle B(0)\rangle$, yielding:

\begin{align}
  \Gamma_\phi &= \frac{g^2\frac{\kappa}{2}}{\Delta^2+(\frac{\kappa}{2})^2}+\frac{\eps_d^2\chi^2\frac{\kappa}{2}}{(\Delta^2+(\frac{\kappa}{2})^2)^2}+\frac{2 g\chi\eps_d\frac{\kappa}{2}\Delta}{(\Delta^2+(\frac{\kappa}{2})^2)^2}\,.
\end{align}

Similar to the $g$ term in Eq.~2, the third term changes sign with detuning, leading to an asymmetry in the dephasing rate around resonance, but in contrast to Eq.~2, this effect requires both a dispersive and longitudinal interaction with the resonator. This asymmetry is observed in the data.  \figref{fig5}{a},{\color{blue}b} show the maximum amplitude extracted for each drive frequency and for two different drive powers. Using Eq.~4, we model the amplitude decay, $A = A_0 e^{-\Gamma_\phi(\tau/2+\delta t)}$, using qubit-resonator coupling strengths and resonator parameters extracted from the phase-shift model that we presented in Eq.~2, which is an independent theory. We extract the maximum amplitude of the exchange oscillations presented in \figref{fig3}{} and plot them with the result obtained from the model in Eq.~4; we find that they show excellent agreement. The asymmetry is present when we have both longitudinal and dispersive effects (\figref{fig5}{a}). The asymmetry is proportional to the last term in Eq.~4, in contrast to the first two terms, which only scale with $g$ and $\chi$, respectively, demonstrating the significant contribution of the longitudinal coupling term to the dynamics of the spin-resonator system.

\section{Towards two-qubit coupling}
We have demonstrated that it is possible to produce a longitudinal coupling to the resonator that exceeds the dispersive term in strength.  However, the resonator decay time is the limiting factor in the current device, and we were not able to generate a two-qubit coupling. We now describe the reasons for this limitation, and consider several paths forward for generating two-qubit coupling by improving system characteristics

\begin{figure}[t]
	\includegraphics[width= 3.4 in]{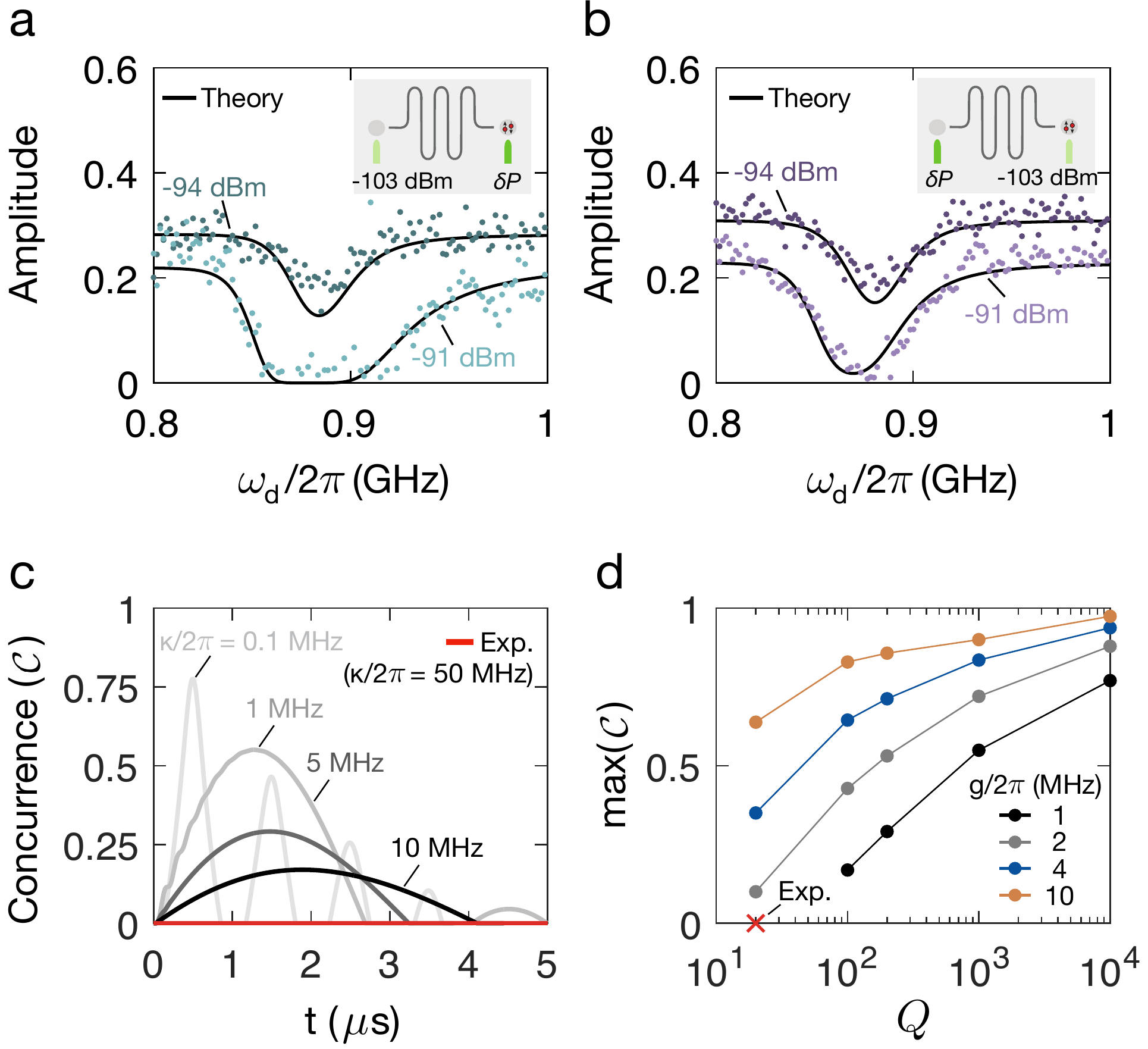}
	\caption{\textbf{Qubit-resonator entanglement and paths towards two-qubit coupling}. Qubit dephasing is analyzed by considering its coupling to photon noise through entanglement with the resonator, also known as measurement-induced dephasing. \textbf{a), b)} Maximum amplitude of exchange oscillations as a function of $\omega_d$ for two driving cases, illustrated in the figure insets.  The third term shows behaviour similar to the. When the right drive is varied (a), an asymmetry is observed in the qubit dephasing, which is captured by the model presented in Eq.~3. Parameters extracted from the phase shift model in Eq.~2 are used in the model instead of fitting the data. When the left drive is varied (b), the qubit dephasing is symmetric and well captured by the model. \textbf{c)} The concurrence is extracted for a set of resonator decay rates. Improving the resonator lifetime greatly enhances two-qubit coupling. \textbf{d)} Maximum concurrence as a function of resonator $Q$ for a set of coupling strengths shows a finite two-qubit entanglement is achieved at $Q=20$ by increasing $g/2\pi$ to $2~\text{MHz}$, a doubling of the experimentally-achieved coupling strength.
	}
	\label{fig5}
\end{figure}

The two-qubit gate proposed in Ref.~\cite{Harvey2018} requires $g\gg \chi$. Since $g$ is proportional to the drive amplitude, the coupling strength can be tuned to reach a regime where the longitudinal coupling dominates, as discussed previously (see \figref{fig4}{b}), by increasing the drive such that $\epsilon_d \gg c_r V_0$. In this regime, where the dispersive term proportional to $\chi$ is negligible, two-qubit coupling can be generated by placing two qubits at opposite antinodes of the resonator, each longitudinally coupled to it. This leads to the following two-qubit Hamiltonian \cite{Harvey2018}
\begin{align}
    H_{12}/\hbar &= \Delta \ad a + g_1(a+\ad)\sigma_{z1}  + g_2(a+\ad)\sigma_{z2}. 
\end{align}
A two-qubit gate is generated by driving the system slightly detuned from the resonator frequency. This allows one to make a closed loop in phase space that corresponds to an accumulated relative phase $\Phi_{12}=\frac{g_1g_2}{2\hbar^2\Delta}t_g$. When $\Phi_{12}=\pi/4$, a CPHASE gate is implemented on the two-qubit system. Disentangling the resonator from the system requires that a full loop is completed, hence setting $\Delta \cdot t_g = 2\pi n$, where $n$ is the number of loops in the resonator phase space. This leads to the optimal detuning $\Delta = 2\sqrt{ng_1g_2}$. Using experimental parameters for the coupling strength $g$, we calculate the entanglement that can be generated by this two-qubit gate (as measured by the concurrence) for several different resonator decay rates by solving the master equation for the two-qubit interaction described by Eq.~5 and including cavity decay. The results are presented in \figref{fig5}{c}, where for each point the optimum detuning is chosen such that it maximizes the concurrence.  In the current device, $\kappa/2\pi \approx 50~\text{MHz}$ giving $Q=\omega_d/\kappa\approx 20$, so it is clear from examination of \figref{fig5}{c} that resonator decay is the main limitation. 
Since resonator decay dominates gate infidelity, increasing the number of loops $n$ improves the two-qubit entanglement as it requires a larger $\Delta$ and thus lower photon number during the gate. However this also increases the gate time, leading to an optimal $n$ set by balancing the resonator decay and limited qubit coherence. For this simulation we assumed typical spin qubit coherence values $T_1 = 100~\mu\text{s}$, $T_2 = 10~\mu\text{s}$ \cite{Dial2013}, and used the corresponding optimal $n$ values. 

While our analysis shows that two-qubit entanglement can be achieved even for modest values of resonator quality factor, the low value of $Q \approx 20$ is currently the main limiting factor in the present device.
Several components in the circuit could be contributing to reducing $Q$ in this experiment. A resonator-bias line was added to give the resonator a DC voltage bias, enabling better tuning of the quantum dot potentials, which allowed both DQDs on the same resonator to be tuned into $\text{\emph{S-T}}_0$ qubits simultaneously \cite{Frey2012, Mi2017}.  With the resonator being very close to each DQD, this can lead to warping of the potential landscape the DQDs are sitting at, making tuning of both qubits challenging. The bias line helps to offset the influence of the resonator, enabling improved control of tuning. This has previously been a major challenge in the research towards performing two-qubit entanglement through a resonator in spin qubits. However, imperfections in the design of the bias line could affect resonator $Q$ (details are described in Methods). In addition, each qubit's DC gates at the voltage antinodes capacitively couple to the resonator, potentially creating  photon leakage paths. These could be further decoupled by implementing LC filters \cite{Frey2012, Mi2017, harveyCollard2020}. Also, coupling to the lossy substrate could induce surface loss, and chemical processing of the surface could lead to surface roughness, both of which reduce the $Q$. These issues would be mitigated by moving to pristine and low-loss silicon substrates, which can be implemented either with Si-Ge based spin qubits or by using flip-chip methods. 

Additionally, two-qubit entanglement could be improved by increasing the longitudinal coupling $g$, as illustrated in \figref{fig5}{d}.  By doubling the coupling strength relative to what was achieved in this experiment, a finite entanglement can be obtained with the current resonator decay time. Further increasing to $g/2\pi = 10~\text{MHz}$ would significantly enhance the two-qubit coupling, achieving concurrence above $50\%$ at lowest $Q$ of 20 and above $90\%$ for values of resonator $Q$ above 1000, already obtained in other silicon-based spin qubit systems \cite{Mi2017,Borjans2020,Samkharadze1123}. Increasing $g$ can be done, for instance, by switching to other superconducting materials with higher kinetic inductance, such as granular aluminum, which has $L_K\approx1-3~\text{nH}/\square$ \cite{Maleeva2018,Kamenov2020}, two orders of magnitude higher than NbN. This would greatly enhance the impedance of the resonator and induce larger voltage fluctuations at the resonator's antinode proximal to the qubit without increasing the dephasing directly.

In conclusion, we have presented a spin qubit-resonator system with a tunable longitudinal coupling, and shown that we are able to enter a strong longitudinal coupling regime. Our measurement sequence allows the qubit-resonator dynamics to be investigated in multiple ways, revealing pronounced effects due to the longitudinal coupling and allowing us to extract the individual coupling strengths of the longitudinal and dispersive terms. This is the first experimental demonstration of a longitudinal coupling between a semiconductor-based spin qubit and superconducting resonator to our knowledge.  Although a $\text{\emph{S-T}}_0$ qubit was used here, it is possible to use any spin qubit provided it has gateable, charge-like states. The results are therefore important steps towards generating a two-qubit coupling through longitudinal coupling and project promising results for a two-qubit gate fidelity by improving resonator parameters such as decay rates by moving to low-loss silicon substrates or increasing the impedance by utilizing higher kinetic inductance materials.

\section*{Methods}
\subsection*{Resonator Fabrication}

The 2DEG was gently removed using a chemical etching process that involves diluted phosphoric etch in $H_2 O$ and $H_2O_2$. Phosphoric acid etches GaAs isotropically leaving the angle of the edge at 45 degrees, for a smooth climb of the resonator to the double quantum dot (DQD). Using sputtering techniques, a thin film of NbN is deposited in the etched area between the two mesas, each of which hosts a DQD. The resonator's climb of the mesa is performed in a second fabrication step, connecting the portion of the resonator residing on the etched part of the sample and the part that capacitively couples to the DQDs on the mesas, thus reducing the potential for resonator disconnects during the climb. The resonator is fabricated with a DC voltage bias line which is placed at the voltage node and should therefore not affect the $Q$ of the resonator. However, the bias line could provide a leakage path for photons in the resonator. To prevent this, the bias line was designed with an LC filter, adapting designs previously described in \cite{Frey2012, Mi2017}.
 
\subsection*{Estimating the drive}

The total attenuation in the measurement setup is estimated as follows: 33 dB attenuation has been added to the coax line inside our dilution fridge with 20 dB on the 4K plate, 10 on the 100 mK plate and 3 dB on the mixing chamber plate. We use a Hittite RF generator that is connected to an IQ circuit with a total attenuation of 25 dB. This goes to a splitter on the output of an Arbitrary Waveform Generator (AWG 5014C). The splitter has an internal loss of 6 dB and an additional 16 dB of attenuation was added to the output port before entering the fridge. The resonator and qubit are driven using one of the RF gates. Using the Sonnet software package to simulate the gate geometry with the resistivity of Ti/Au, we further include 20 dB of loss due to the RF gate itself. Altogether, we find the total attenuation of the circuit to be $-$100 dB.

Several values of the drive power are used in the experiment to drive the qubit-resonator system into different regimes. As an example, consider the total power used the experiment of $-$97 dBm converted to 1.58$\cdot 10^{-13}$ Watts. Then using the relation between power and number of photons \cite{Clerk2010}:
\begin{align}
  \langle n \rangle = 4P Q^2/(\hbar\omega_r^2Q_c)\,. 
\end{align}
The coupling quality factor, $Q_c$, is estimated from a simulation of the device structure, which gives $Q_c = 31,457$. An estimate of the total loaded $Q$ is found from experimental values $Q = \omega_r/\kappa = 18.72$. This gives an estimate for the average number of photons in the resonator for a given drive power to be $\langle n\rangle = 2.19 $. From $\langle n\rangle$ and $\kappa$ we find the drive amplitude is given by $\eps_d/2\pi = \sqrt{\langle n\rangle}\kappa/2=34.74 \text{ MHz}$.

\subsection*{Calculating power spectral densities}

We compute the power spectral densities used in the dephasing model described by Eq.~(3) of the main text, where we consider the limit $g,\chi \ll \kappa$. The dephasing is attributed to photon noise, or fluctuations in the photon number operator $n = a^\dagger a$ and $x = (a+ a^\dagger)$ that couples to $\sigma_z$ with a coupling strength $\chi$ and $g$ respectively. Additional terms arise from cross couplings, which will be computed here as well.
Calculations of the power spectral densities, $S_{AB}(\omega)=\int_{-\infty}^{\infty}dt C_{AB}(t)e^{-i\omega t}$, for each dephasing term in Eq.~(3) are carried out by first computing the correlation functions, according to the general definition 
\begin{align}
 C_{AB}(t) = \langle A(t)B(0)\rangle - \langle A(t)\rangle\langle B(0)\rangle \,.
\end{align}
Then, we move into the displacement frame $d = a-\beta$ ($d^{\dagger} = a^{\dagger}-\beta^{\star}$) where we define a coherent state $ \beta $ such that the cavity equilibrium state is $\ket{0}$. We calculate for $A=B=x$ and find that the dependence on $\beta$ cancels, yielding 
\begin{align}
 C^g_{xx}(t) = \langle d(t)d^{\dagger}(0)\rangle\ = e^{-i\Delta t -\kappa/2 t}\,.
\end{align}
The power spectral density for this term then takes the form
\begin{align}
 S^g_{xx}(\omega) = \frac{\kappa}{(\omega-\Delta)^2+(\frac{\kappa}{2})^2}\,.
\end{align}
Similarly, for the correlation function with $A=B=n$, 
we have similar simplifications except for a term that scales with the mean number of photons $\bar{n}= \beta^{\star}\beta$.  The correlation function can be written as  
\begin{align}
 C_{nn}(t) = \bar{n}\langle d(t)d^\dagger(0)\rangle = \bar{n}e^{-i\Delta t -\kappa/2 t}\,.
\end{align}
The power spectral density for this second term takes the form
\begin{align}
 S^g_{nn}(\omega) = \frac{\bar{n}\kappa}{(\omega-\Delta)^2+(\frac{\kappa}{2})^2}\,.
\end{align}
Finally, the cross coupling terms are computed, where we consider both $A=n$, $B=x$ as well as $A=x$, $B=n$.
We find 
\begin{align}
C_{nx}(t)&=\beta^{\star}\langle d(t)d^\dagger(0)\rangle \,, \\
C_{xn}(t)&= \beta\langle d(t)d^\dagger(0)\rangle\,.
\end{align}
Combining both gives
\begin{align}
C_{nx}(t) + C_{xn}(t) = 2{\rm Re}(\beta)e^{-i\Delta t -\kappa/2 t}\,,
\end{align}
and the power spectral density becomes
\begin{align}
 S^{g\chi}(\omega) = \frac{2{\rm Re}(\beta)\kappa}{(\omega-\Delta)^2+(\frac{\kappa}{2})^2}\,.
\end{align}
Note that this term scales with $\sqrt{\bar n}$. 

As a final step, we wish to express these three power spectral densities in terms of the drive $\eps_d$.  Using $\beta=\frac{\eps_d}{\Delta+i\frac{\kappa}{2}}$ we find
\begin{align}
2{\rm Re}(\beta)= \frac{2\eps_d\Delta}{\Delta^2+(\frac{\kappa}{2})^2} \,.
\end{align}
Note, this only holds if $\eps_d$ and $g$ share the same phase. Notice that we now have a linear dependence on detuning in the numerator and this term will change sign with $\Delta$. Similarly, we can express $\bar{n}$ in terms of $\eps_d$ as well:
\begin{align}
\bar{n}=\frac{\eps_d^2}{\Delta^2+(\frac{\kappa}{2})^2} \,.
\end{align}
These three terms for the power spectral densities lead directly to Eq.~(4) in the main text.

\emph{Acknowledgements:} 
We acknowledge useful discussions with and feedback from Y. Liu, L. Orona, A. Pierce and N. Poniatowski. This work is supported by the Quantum Science Center (QSC), a National Quantum Information Science Research Center of the U.S. Department of Energy (DOE), the Gordon and Betty Moore Foundation through Grant GBMF 9468, the National Science Foundation under Grant No. DMR-1708688, and the STC Center for Integrated Quantum Materials, NSF Grant No. DMR-1231319. S.D.B acknowledges support from the Australian Research Council (ARC) project number CE170100009.\\

\emph{Author contributions:} 
C.B. and S.H. performed low-temperature measurements and fabricated the sample. S.F., G.G. and M.M. grew the wafer. C.B., S.B. and U.V. performed the data analysis and modeling. C.B., S.H., U.V and A.Y. designed the experiment. All authors discussed the results and commented on the manuscript.\\

\emph{Data Availability:} 
The data that support the findings of this study are available from the corresponding author upon reasonable request.

\bibliography{Bibliography}

\begin{thebibliography}{10}
\expandafter\ifx\csname url\endcsname\relax
  \def\url#1{\texttt{#1}}\fi
\expandafter\ifx\csname urlprefix\endcsname\relax\def\urlprefix{URL }\fi
\providecommand{\bibinfo}[2]{#2}
\providecommand{\eprint}[2][]{\url{#2}}

\bibitem{Loss1998}
\bibinfo{author}{Loss, D.} \& \bibinfo{author}{DiVincenzo, D.~P.}
\newblock \bibinfo{title}{Quantum computation with quantum dots}.
\newblock \emph{\bibinfo{journal}{Phys. Rev. A}} \textbf{\bibinfo{volume}{57}},
  \bibinfo{pages}{120--126} (\bibinfo{year}{1998}).

\bibitem{Koppens2006}
\bibinfo{author}{Koppens, F. H.~L.} \emph{et~al.}
\newblock \bibinfo{title}{Driven coherent oscillations of a single electron
  spin in a quantum dot}.
\newblock \emph{\bibinfo{journal}{Nature}} \textbf{\bibinfo{volume}{442}},
  \bibinfo{pages}{766--771} (\bibinfo{year}{2006}).

\bibitem{Ladriere2008}
\bibinfo{author}{Pioro-Ladrière, M.} \emph{et~al.}
\newblock \bibinfo{title}{Electrically driven single-electron spin resonance in
  a slanting zeeman field}.
\newblock \emph{\bibinfo{journal}{Nature Physics}}
  \textbf{\bibinfo{volume}{4}}, \bibinfo{pages}{776--779}
  (\bibinfo{year}{2008}).

\bibitem{Kim2014}
\bibinfo{author}{Kim, D.} \emph{et~al.}
\newblock \bibinfo{title}{Quantum control and process tomography of a
  semiconductor quantum dot hybrid qubit}.
\newblock \emph{\bibinfo{journal}{Nature}} \textbf{\bibinfo{volume}{511}},
  \bibinfo{pages}{70--74} (\bibinfo{year}{2014}).

\bibitem{Eng2015}
\bibinfo{author}{Eng, K.} \emph{et~al.}
\newblock \bibinfo{title}{Isotopically enhanced triple-quantum-dot qubit}.
\newblock \emph{\bibinfo{journal}{Science Advances}}
  \textbf{\bibinfo{volume}{1}} (\bibinfo{year}{2015}).

\bibitem{Yang2019}
\bibinfo{author}{Yang, C.~H.} \emph{et~al.}
\newblock \bibinfo{title}{Silicon qubit fidelities approaching incoherent noise
  limits via pulse engineering}.
\newblock \emph{\bibinfo{journal}{Nature Electronics}}
  \textbf{\bibinfo{volume}{2}}, \bibinfo{pages}{151--158}
  (\bibinfo{year}{2019}).

\bibitem{Dohun2014}
\bibinfo{author}{Dohun, K.} \emph{et~al.}
\newblock \bibinfo{title}{Quantum control and process tomography of a
  semiconductor quantum dot hybrid qubit}.
\newblock \emph{\bibinfo{journal}{Nature}} \textbf{\bibinfo{volume}{511}},
  \bibinfo{pages}{70--74} (\bibinfo{year}{2014}).

\bibitem{Medford2014}
\bibinfo{author}{Medford, J.} \emph{et~al.}
\newblock \bibinfo{title}{Self-consistent measurement and state tomography of
  an exchange-only spin qubit}.
\newblock \emph{\bibinfo{journal}{Nature Nanotechnology}}
  \textbf{\bibinfo{volume}{8}}, \bibinfo{pages}{654--659}
  (\bibinfo{year}{2014}).

\bibitem{Wu2014}
\bibinfo{author}{Wu, X.} \emph{et~al.}
\newblock \bibinfo{title}{Two-axis control of a singlet{\textendash}triplet
  qubit with an integrated micromagnet}.
\newblock \emph{\bibinfo{journal}{Proceedings of the National Academy of
  Sciences}} \textbf{\bibinfo{volume}{111}}, \bibinfo{pages}{11938--11942}
  (\bibinfo{year}{2014}).

\bibitem{Cerfontaine2020}
\bibinfo{author}{Cerfontaine, P.} \emph{et~al.}
\newblock \bibinfo{title}{Closed-loop control of a gaas-based singlet-triplet
  spin qubit with 99.5\% gate fidelity and low leakage}.
\newblock \emph{\bibinfo{journal}{Nature Communications}}
  \textbf{\bibinfo{volume}{11}} (\bibinfo{year}{2020}).

\bibitem{Nichol2017}
\bibinfo{author}{Nichol, J.~M.} \emph{et~al.}
\newblock \bibinfo{title}{High-fidelity entangling gate for double-quantum-dot
  spin qubits}.
\newblock \emph{\bibinfo{journal}{npj Quantum Information}}
  \textbf{\bibinfo{volume}{3}} (\bibinfo{year}{2017}).

\bibitem{Fujita}
\bibinfo{author}{Fujita, T.}, \bibinfo{author}{Baart, T.~A.},
  \bibinfo{author}{Reichl, C.}, \bibinfo{author}{Wegscheider, W.} \&
  \bibinfo{author}{Vandersypen, L. M.~K.}
\newblock \bibinfo{title}{Coherent shuttle of electron-spin states}.
\newblock \emph{\bibinfo{journal}{npj Quantum Information}}
  \textbf{\bibinfo{volume}{3}}, \bibinfo{pages}{22} (\bibinfo{year}{2017}).

\bibitem{Mills2019}
\bibinfo{author}{Mills, A.~R.} \emph{et~al.}
\newblock \bibinfo{title}{Shuttling a single charge across a one-dimensional
  array of silicon quantum dots}.
\newblock \emph{\bibinfo{journal}{Nature Communications}}
  \textbf{\bibinfo{volume}{10}} (\bibinfo{year}{2019}).

\bibitem{yoneda2020coherent}
\bibinfo{author}{Yoneda, J.} \emph{et~al.}
\newblock \bibinfo{title}{Coherent spin qubit transport in silicon}.
\newblock \emph{\bibinfo{journal}{arXiv preprint arXiv:2008.04020}}
  (\bibinfo{year}{2020}).

\bibitem{Qiao2021}
\bibinfo{author}{Qiao, H.} \emph{et~al.}
\newblock \bibinfo{title}{Long-distance superexchange between semiconductor
  quantum-dot electron spins}.
\newblock \emph{\bibinfo{journal}{Phys. Rev. Lett.}}
  \textbf{\bibinfo{volume}{126}}, \bibinfo{pages}{017701}
  (\bibinfo{year}{2021}).

\bibitem{Samkharadze1123}
\bibinfo{author}{Samkharadze, N.} \emph{et~al.}
\newblock \bibinfo{title}{Strong spin-photon coupling in silicon}.
\newblock \emph{\bibinfo{journal}{Science}} \textbf{\bibinfo{volume}{359}},
  \bibinfo{pages}{1123--1127} (\bibinfo{year}{2018}).

\bibitem{Landig2018}
\bibinfo{author}{Landig, A.~J.} \emph{et~al.}
\newblock \bibinfo{title}{Coherent spin–photon coupling using a resonant
  exchange qubit}.
\newblock \emph{\bibinfo{journal}{Nature}} \textbf{\bibinfo{volume}{560}},
  \bibinfo{pages}{179--184} (\bibinfo{year}{2018}).

\bibitem{Borjans2020}
\bibinfo{author}{Borjans, F.} \emph{et~al.}
\newblock \bibinfo{title}{Split-gate cavity coupler for silicon circuit quantum
  electrodynamics}.
\newblock \emph{\bibinfo{journal}{Applied Physics Letters}}
  \textbf{\bibinfo{volume}{116}}, \bibinfo{pages}{234001}
  (\bibinfo{year}{2020}).

\bibitem{Mi2017}
\bibinfo{author}{Mi, X.}, \bibinfo{author}{Cady, J.~V.},
  \bibinfo{author}{Zajac, D.~M.}, \bibinfo{author}{Deelman, P.~W.} \&
  \bibinfo{author}{Petta, J.~R.}
\newblock \bibinfo{title}{Strong coupling of a single electron in silicon to a
  microwave photon}.
\newblock \emph{\bibinfo{journal}{Science}} \textbf{\bibinfo{volume}{355}},
  \bibinfo{pages}{156--158} (\bibinfo{year}{2017}).

\bibitem{Mi20172}
\bibinfo{author}{Mi, X.} \emph{et~al.}
\newblock \bibinfo{title}{Circuit quantum electrodynamics architecture for
  gate-defined quantum dots in silicon}.
\newblock \emph{\bibinfo{journal}{Applied Physics Letters}}
  \textbf{\bibinfo{volume}{110}}, \bibinfo{pages}{043502}
  (\bibinfo{year}{2017}).

\bibitem{Borjans20202}
\bibinfo{author}{Borjans, F.}, \bibinfo{author}{Croot, X.~G.},
  \bibinfo{author}{Mi, X.}, \bibinfo{author}{Gullans, M.~J.} \&
  \bibinfo{author}{Petta, J.~R.}
\newblock \bibinfo{title}{Resonant microwave-mediated interactions between
  distant electron spins}.
\newblock \emph{\bibinfo{journal}{Nature}} \textbf{\bibinfo{volume}{577}},
  \bibinfo{pages}{195--198} (\bibinfo{year}{2020}).

\bibitem{Wallraff2004}
\bibinfo{author}{Wallraff, A.} \emph{et~al.}
\newblock \bibinfo{title}{Strong coupling of a single photon to a
  superconducting qubit using circuit quantum electrodynamics}.
\newblock \emph{\bibinfo{journal}{Nature}} \textbf{\bibinfo{volume}{431}},
  \bibinfo{pages}{162--167} (\bibinfo{year}{2004}).

\bibitem{Blais2004}
\bibinfo{author}{Blais, A.}, \bibinfo{author}{Huang, R.-S.},
  \bibinfo{author}{Wallraff, A.}, \bibinfo{author}{Girvin, S.~M.} \&
  \bibinfo{author}{Schoelkopf, R.~J.}
\newblock \bibinfo{title}{Cavity quantum electrodynamics for superconducting
  electrical circuits: An architecture for quantum computation}.
\newblock \emph{\bibinfo{journal}{Phys. Rev. A}} \textbf{\bibinfo{volume}{69}},
  \bibinfo{pages}{062320} (\bibinfo{year}{2004}).

\bibitem{Kerman2013}
\bibinfo{author}{Kerman, A.~J.}
\newblock \bibinfo{title}{Quantum information processing using quasiclassical
  electromagnetic interactions between qubits and electrical resonators}.
\newblock \emph{\bibinfo{journal}{New Journal of Physics}}
  \textbf{\bibinfo{volume}{15}}, \bibinfo{pages}{123011}
  (\bibinfo{year}{2013}).

\bibitem{Schuetz2017}
\bibinfo{author}{Schuetz, M. J.~A.}, \bibinfo{author}{Giedke, G.},
  \bibinfo{author}{Vandersypen, L. M.~K.} \& \bibinfo{author}{Cirac, J.~I.}
\newblock \bibinfo{title}{High-fidelity hot gates for generic spin-resonator
  systems}.
\newblock \emph{\bibinfo{journal}{Phys. Rev. A}} \textbf{\bibinfo{volume}{95}},
  \bibinfo{pages}{052335} (\bibinfo{year}{2017}).

\bibitem{Billangeon2015}
\bibinfo{author}{Billangeon, P.-M.}, \bibinfo{author}{Tsai, J.~S.} \&
  \bibinfo{author}{Nakamura, Y.}
\newblock \bibinfo{title}{Circuit-qed-based scalable architectures for quantum
  information processing with superconducting qubits}.
\newblock \emph{\bibinfo{journal}{Phys. Rev. B}} \textbf{\bibinfo{volume}{91}},
  \bibinfo{pages}{094517} (\bibinfo{year}{2015}).

\bibitem{Didier2015}
\bibinfo{author}{Didier, N.}, \bibinfo{author}{Bourassa, J.} \&
  \bibinfo{author}{Blais, A.}
\newblock \bibinfo{title}{Fast quantum nondemolition readout by parametric
  modulation of longitudinal qubit-oscillator interaction}.
\newblock \emph{\bibinfo{journal}{Phys. Rev. Lett.}}
  \textbf{\bibinfo{volume}{115}}, \bibinfo{pages}{203601}
  (\bibinfo{year}{2015}).

\bibitem{Royer2017}
\bibinfo{author}{Royer, B.}, \bibinfo{author}{Grimsmo, A.~L.},
  \bibinfo{author}{Didier, N.} \& \bibinfo{author}{Blais, A.}
\newblock \bibinfo{title}{Fast and high-fidelity entangling gate through
  parametrically modulated longitudinal coupling}.
\newblock \emph{\bibinfo{journal}{{Quantum}}} \textbf{\bibinfo{volume}{1}},
  \bibinfo{pages}{11} (\bibinfo{year}{2017}).

\bibitem{Ruskov2021}
\bibinfo{author}{Ruskov, R.} \& \bibinfo{author}{Tahan, C.}
\newblock \bibinfo{title}{Modulated longitudinal gates on encoded spin qubits
  via curvature couplings to a superconducting cavity}.
\newblock \emph{\bibinfo{journal}{Phys. Rev. B}}
  \textbf{\bibinfo{volume}{103}}, \bibinfo{pages}{035301}
  (\bibinfo{year}{2021}).

\bibitem{Harvey2018}
\bibinfo{author}{Harvey, S.~P.} \emph{et~al.}
\newblock \bibinfo{title}{Coupling two spin qubits with a high-impedance
  resonator}.
\newblock \emph{\bibinfo{journal}{Phys. Rev. B}} \textbf{\bibinfo{volume}{97}},
  \bibinfo{pages}{235409} (\bibinfo{year}{2018}).

\bibitem{Molmer1999}
\bibinfo{author}{M\o{}lmer, K.} \& \bibinfo{author}{S\o{}rensen, A.}
\newblock \bibinfo{title}{Multiparticle entanglement of hot trapped ions}.
\newblock \emph{\bibinfo{journal}{Phys. Rev. Lett.}}
  \textbf{\bibinfo{volume}{82}}, \bibinfo{pages}{1835--1838}
  (\bibinfo{year}{1999}).

\bibitem{Kirchmair2009}
\bibinfo{author}{Kirchmair, G.} \emph{et~al.}
\newblock \bibinfo{title}{Deterministic entanglement of ions in thermal states
  of motion}.
\newblock \emph{\bibinfo{journal}{New Journal of Physics}}
  \textbf{\bibinfo{volume}{11}}, \bibinfo{pages}{023002}
  (\bibinfo{year}{2009}).

\bibitem{Haffner2008}
\bibinfo{author}{Häffner, H.}, \bibinfo{author}{Roos, C.} \&
  \bibinfo{author}{Blatt, R.}
\newblock \bibinfo{title}{Quantum computing with trapped ions}.
\newblock \emph{\bibinfo{journal}{Physics Reports}}
  \textbf{\bibinfo{volume}{469}}, \bibinfo{pages}{155--203}
  (\bibinfo{year}{2008}).

\bibitem{Tinkham}
\bibinfo{author}{Tinkham, M.}
\newblock \emph{\bibinfo{title}{Introduction to Superconductivity}}
  (\bibinfo{publisher}{Dover Books on Physics}, \bibinfo{year}{2004}).

\bibitem{Taylor2008}
\bibinfo{author}{Taylor, J.~M.} \emph{et~al.}
\newblock \bibinfo{title}{{High-sensitivity diamond magnetometer with nanoscale
  resolution}}.
\newblock \emph{\bibinfo{journal}{Nature Physics}}
  \textbf{\bibinfo{volume}{4}}, \bibinfo{pages}{810--816}
  (\bibinfo{year}{2008}).

\bibitem{Medford2012}
\bibinfo{author}{Medford, J.} \emph{et~al.}
\newblock \bibinfo{title}{Scaling of dynamical decoupling for spin qubits}.
\newblock \emph{\bibinfo{journal}{Phys. Rev. Lett.}}
  \textbf{\bibinfo{volume}{108}}, \bibinfo{pages}{086802}
  (\bibinfo{year}{2012}).

\bibitem{Dial2013}
\bibinfo{author}{Dial, O.~E.} \emph{et~al.}
\newblock \bibinfo{title}{Charge noise spectroscopy using coherent exchange
  oscillations in a singlet-triplet qubit}.
\newblock \emph{\bibinfo{journal}{Phys. Rev. Lett.}}
  \textbf{\bibinfo{volume}{110}}, \bibinfo{pages}{146804}
  (\bibinfo{year}{2013}).

\bibitem{Clerk2010}
\bibinfo{author}{Clerk, A.~A.}, \bibinfo{author}{Devoret, M.~H.},
  \bibinfo{author}{Girvin, S.~M.}, \bibinfo{author}{Marquardt, F.} \&
  \bibinfo{author}{Schoelkopf, R.~J.}
\newblock \bibinfo{title}{Introduction to quantum noise, measurement, and
  amplification}.
\newblock \emph{\bibinfo{journal}{Rev. Mod. Phys.}}
  \textbf{\bibinfo{volume}{82}}, \bibinfo{pages}{1155--1208}
  (\bibinfo{year}{2010}).

\bibitem{Gambetta2006}
\bibinfo{author}{Gambetta, J.} \emph{et~al.}
\newblock \bibinfo{title}{Qubit-photon interactions in a cavity:
  Measurement-induced dephasing and number splitting}.
\newblock \emph{\bibinfo{journal}{Phys. Rev. A}} \textbf{\bibinfo{volume}{74}},
  \bibinfo{pages}{042318} (\bibinfo{year}{2006}).

\bibitem{Ithier2005}
\bibinfo{author}{Ithier, G.} \emph{et~al.}
\newblock \bibinfo{title}{Decoherence in a superconducting quantum bit
  circuit}.
\newblock \emph{\bibinfo{journal}{Phys. Rev. B}} \textbf{\bibinfo{volume}{72}},
  \bibinfo{pages}{134519} (\bibinfo{year}{2005}).

\bibitem{Frey2012}
\bibinfo{author}{Frey, T.} \emph{et~al.}
\newblock \bibinfo{title}{Dipole coupling of a double quantum dot to a
  microwave resonator}.
\newblock \emph{\bibinfo{journal}{Phys. Rev. Lett.}}
  \textbf{\bibinfo{volume}{108}}, \bibinfo{pages}{046807}
  (\bibinfo{year}{2012}).

\bibitem{harveyCollard2020}
\bibinfo{author}{Harvey-Collard, P.} \emph{et~al.}
\newblock \bibinfo{title}{On-chip microwave filters for high-impedance
  resonators with gate-defined quantum dots}.
\newblock \emph{\bibinfo{journal}{Phys. Rev. Applied}}
  \textbf{\bibinfo{volume}{14}}, \bibinfo{pages}{034025}
  (\bibinfo{year}{2020}).

\bibitem{Maleeva2018}
\bibinfo{author}{Maleeva, N.} \emph{et~al.}
\newblock \bibinfo{title}{Circuit quantum electrodynamics of granular aluminum
  resonators}.
\newblock \emph{\bibinfo{journal}{Nature Communications}}
  \textbf{\bibinfo{volume}{9}} (\bibinfo{year}{2018}).

\bibitem{Kamenov2020}
\bibinfo{author}{Kamenov, P.} \emph{et~al.}
\newblock \bibinfo{title}{Granular aluminum meandered superinductors for
  quantum circuits}.
\newblock \emph{\bibinfo{journal}{Phys. Rev. Applied}}
  \textbf{\bibinfo{volume}{13}}, \bibinfo{pages}{054051}
  (\bibinfo{year}{2020}).

\end{thebibliography}

\end{document}